\documentclass{aastex62}

\received{February 20, 2020}
\revised{November 4, 2020}
\accepted{November 4, 2020}
\submitjournal{ApJ} 

\shorttitle{Goodarzi et al.}
\shortauthors{Goodarzi et al.}

\begin{document}

\title{Flare Activity and Magnetic Feature Analysis of the Flare Stars II: Sub-Giant Branch}

\correspondingauthor{Hadis Goodarzi}
\email{goodarzi@ipm.ir, hadisgoodarzy@yahoo.com}

\author[0000-0001-9128-2012]{Hadis Goodarzi}
\affiliation{School of Astronomy, Institute for Research in Fundamental Sciences (IPM) \\
P.O. Box 19395-5746, Tehran, Iran}

\author{Ahmad Mehrabi}
\affiliation{Department of Physics, Bu Ali Sina University \\
65178, 016016, Hamedan, Iran}

\affiliation{School of Astronomy, Institute for Research in Fundamental Sciences (IPM) \\
P.O. Box 19395-5746, Tehran, Iran}

\author{Habib G.Khosroshahi}
\affiliation{School of Astronomy, Institute for Research in Fundamental Sciences (IPM) \\
P.O. Box 19395-5746, Tehran, Iran}


\author{Han He}
\affiliation{CAS Key Laboratory of Solar Activity \\ 
National Astronomical Observatories, Chinese Academy of Sciences \\
Beijing, People's Republic of China}

\begin{abstract}
We present an investigation of the magnetic activity and flare characteristics of the sub-giant stars mostly from F and G spectral types and compare the results with the main-sequence (MS) stars. The light curve of 352 stars on the sub-giant branch (SGB) from the Kepler mission is analyzed in order to infer stability, relative coverage and contrast of the magnetic structures and also flare properties using three flare indexes. The results show that: (i)  Relative coverage and contrast of the magnetic features along with rate, power and magnitude of flares increase on the SGB due to the deepening of the convective zone and more vigorous magnetic field production (ii) Magnetic activity of the F and G-type stars on the SGB does not show dependency to the rotation rate and does not obey the saturation regime. This is the opposite of what we saw for the main sequence, in which the G-, K- and M-type stars show clear dependency to the Rossby number; (iii) The positive relationship between the magnetic features stability and their relative coverage and contrast remains true on the SGB, though it has lower dependency coefficient in comparison with the MS; (iv) Magnetic proxies and flare indexes of the SGB stars increase with increasing the relative mass of the convective zone.    

\end{abstract}

\section{Introduction} \label{sec:intro}
 
The study of stellar magnetic activity is of primary importance to better understand the formation and evolution of the stars. Flare activity can have physical and chemical impacts on the early phase of the star and planet formation because its ionizing radiation affects the degree of ionization in accretion disk of the young stellar objects \citep{Benz10,Herbst&Klemperer73,Dalgarno&Black76,Herbst&van Dishoeck09}. 
Stellar brightness modulation and flare events are the most outstanding magnetic activity indicators that can be inferred from the observed light curves. 

Kepler mission has made it possible to study the light-curve of more than 195000 targets. The initial purpose of this mission was to detect the earth like planets in the habitable zone of stars \citep{Borucki2010,Koch2010}. However, its long duration and continuous observations, along with high precision photometry, provides the possibility of studying the magnetic and flare activity for a large sample of stars \citep{Basri10,Basri11,Chaplin11,Frasca11,Kurtz11,Walkowicz2011}. Analyzing the observations of a large number of stars with the same instrumental setup, data reduction and observing techniques decreases the random errors and increases the robustness of the results \citep{Jenkins2011}; this is one of the great benefits of the data provided by the Kepler mission.
  
In the stellar dynamo model, surface activity of stars is the result of interaction between the magnetic field, rotation and convection itself or a thin shear zone between the convective and radiative zone \citep{Parker55,Parker70,Steenbeck&Krause69,Moffat78,Gilman1983,Glatzmaier84,Spruit2002}, so the strength of produced magnetic field depends on the rotation rate and the volume of convective envelope \citep{Walter81,Vilhu84}.

Magnetic activity studies on the samples of main-sequence stars \citep{Vaughan81,Galloway&Weiss81,Mangeney&Praderie84,Noyes84,Vilhu84,Simon85,Middelkoop82} confirms that the magnetic activity has positive correlation with the rotation rate, i.e., by decreasing rotation period, magnetic activity increases. 
However, at high rotation rates, most of the activity indicators saturate and do not increase further \citep{Noyes84,Vilhu84,Vilhu&Walter87}.
This saturation regime has been also verified using spectroscopic data and magnetic field sensitive lines; however, rotational line broadening in fast rotators prevents measuring the magnetic flux in the regime of saturated activity Reiners2009.

In the subgiant phase, Helium burning has been ceased in the core and commenced in a shell outside the core. Thus, the star is brighter than a normal main-sequence star of the same spectral class, but not as bright as the giant stars \citep{Adams1935}. The core of the star contracts and its envelope expands when the star leave the main sequence (MS) to the subgiant branch (SGB), so the moment of inertia increases. Furthermore, as even hot rapidly rotating stars ($>$6200K) develop convective envelope on the SGB, wind-driven angular momentum loss can occur for both fast and slow rotating stars when they leave the MS \citep{Hoyle60,Schatzman62,Gray89,Schrijver&Pols93,Van Saders2013}. 

Up to now, most of the statistical activity studies have been done on the main sequence and giant stars  \citep{Haisch&Simon82,Fekel&Balachandran93,Dempsey93,Notsu2013,Shibayama2013,He2015,Mehrabi2017} and less attention has been paid to the subgiant branch as the intermediate evolutionary state of the stars between dwarfs and giants \citep{Stromberg30,Sandage2003}, while considering magnetic activity and other phenomena in giant and subgiant stars may help researchers to better understand the mechanism of such events on the main sequence stars \citep{Katsova2018}. For ex. \cite{Lambert&Ries81} studied Carbon, Nitrogen and Oxygen abundance of 32 giants and subgiants from G and K spectral types. By evaluating of the amount of processed material dredging to the surface of these stars, they concluded that mixing occurs in the radiative zone - between the core and the outer convective zone - of the main sequence stars.

However, when comparing the young and evolved stars with each other or extending their results, different preliminary conditions ot them should be taken into account. For ex. \cite{Judge86} analyzed emission lines formed in the outer atmosphere of late-type giant stars. He found that these emission lines of low-gravity stars are formed in lower temperatures than those at which they would be formed in a solar-type chromospheric and transition region. Also, by analyzing X-ray and ultraviolet observations of 91 targets from spectral type F-K, \cite{Ayres95} found that coronal activity becomes progressively more sensitive to the spectral types when the star leaves the MS.

\cite{Ayres&Linsky80} studied Capella system consisting of three similar giants stars (we now know that it is a quadruple system), they concluded that the origin of the extraordinary high -chromospheric and transition region-activity level of the late-F secondary is due to its higher rotation rate. \cite{Middelkoop&Zwaan81} used \cite{Wilson76} eye estimation of the CaII H and K emission intensities for 500 evolved stars. They concluded that rapidly rotating G-type giants tend to show enhanced CaII H and K emissions and suggested that this is probably true among other spectral types of giants and subgiants.

\cite{Simon&Drake89} analyzed a sample of cool giant and subgiant stars by focusing on their C II 1335A and C IV 1550A emission lines forming in high chromosphere and transition region, respectively. They found strong correlation between the activity and rotation in the stars later than F5, however, emission of early F stars is independent of the rotation rate.
\cite{Garcia-Alvarez06} analyzed high resolution Chandra X-ray spectra of late-type giants in different main stages of their lifetime, they suggested that the structural changes during the evolution of late type giants could be the reason of observed differences in coronal abundances and temperature structure. Particularly, the size of the convective zone coupled with the rotation rate has the dominant role in coronal characteristics.
  
At first, the number of subgiants was underestimated in the initial Kepler Input Catalog (KIC) and the KSPC revisions \citep{Huber2011,Huber14} because the surface gravity determination was carried out by broadband photometry which is not accurate enough. Later, other calculations of surface gravity by short-timescale brightness variations or "flicker" showed that nearly 50\% of the bright planet-host stars are subgiants \citep{Bastien2014}. 
By combining Gaia Data Release 2 parallaxes with the Kepler stellar properties Catalog, \cite{Berger2018} reported the revised radii and evolutionary state of 177,911 Kepler stars. They found that about 67\% (120000) of all Kepler targets are main-sequence stars, 21\% (37000) subgiants, and 12\% (21000) red giants.

In our previous paper (\cite{Goodarzi2019}, hereafter paper I) we analyzed 1740 main-sequence stars from F, G, K and M spectral types and studied relationship between the magnetic feature characteristics and the identified flare activity.
In this paper, we focus on the magnetic properties and flare activity of the subgiant stars and compare the results with the main-sequence. In section \ref{sec:process}, we explain our data selection, then, magnetic proxies and flare indexes are described in section \ref{sec:param}. Results and discussion are presented in section \ref{sec:results}, and finally, we conclude in section \ref{sec:conclusion}.

\section{Data Selection and processing} \label{sec:process}

We select subgiant stars from the revised radii and evolutionary state catalog by \cite{Berger2018}. We choose the targets that their flare activity was confirmed in either of the catalogs by \cite{Davenport2016} or \cite{VanDoorsselaere2017}, and their rotation period was determined in the catalog by \cite{McQuillan2014}. We also remove all the binary candidates reported by \cite{Berger2018}. Final sample contains 352 single-, flare-, subgiant stars (101 F-, 234 G- and 17 K-type). Long-cadence data from quarters Q2-Q16 (data release 25) are selected for analysis and other quarters are neglected because of their short duration. 
 
\begin{figure}
	\gridline{\fig{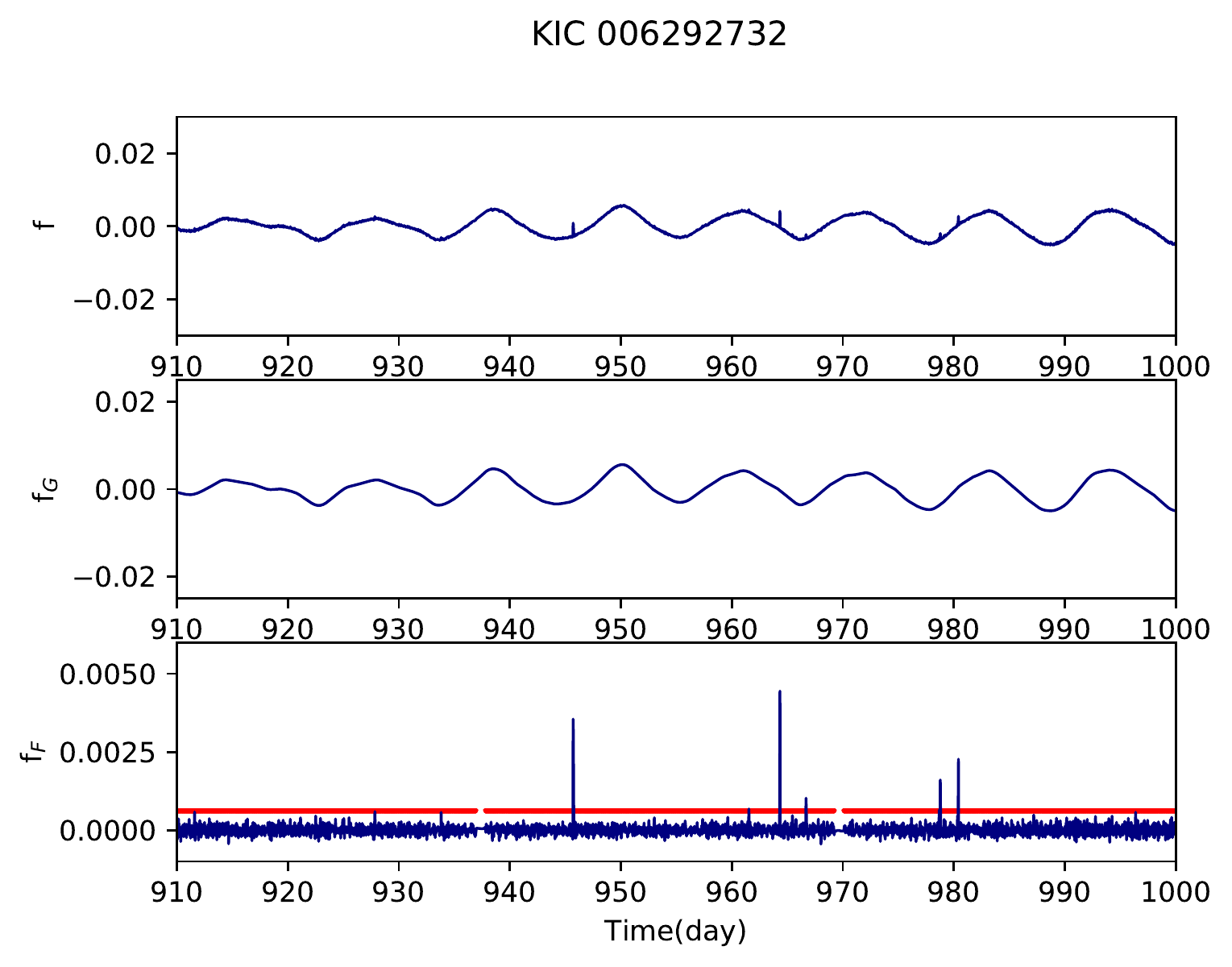}{0.51\textwidth}{(a)}
		\fig{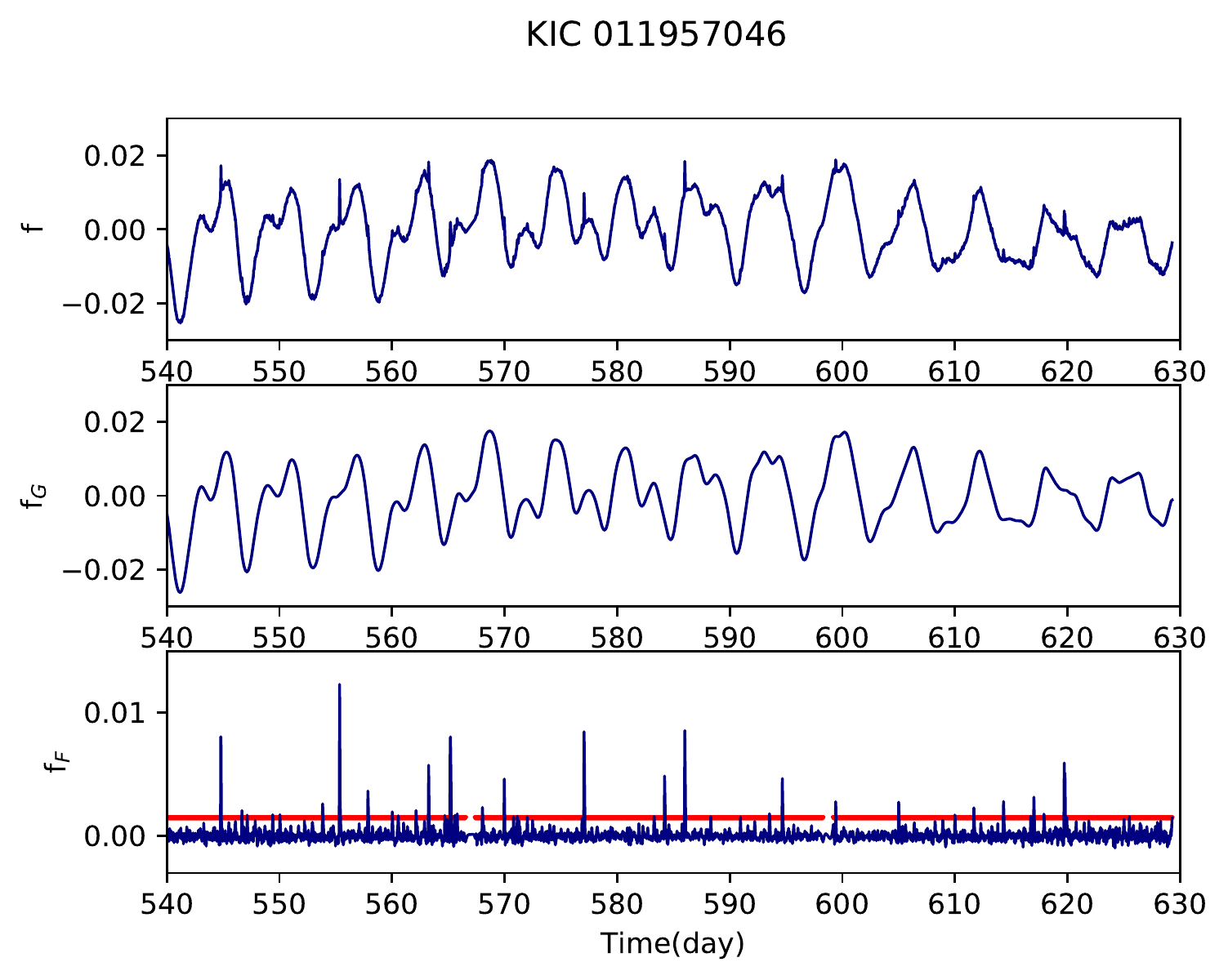}{0.51\textwidth}{(b)}}
	\caption{The light curve of two Kepler targets (KIC 6292732, left) and (KIC 11957046, right) with different magnitude of light curve variations and hence magnetic activity levels. Top panels show the normalized flux (f), the middle panels are normalized gradual component ($f_{G}$) and bottom panels are normalized flare component ($f_{F}$). The height of red horizontal lines in the bottom panels shows threshold value for flare detection and its length shows the valid observation time, the gaps are due to the spacecraft data down-link.}
	\label{Fig1}
\end{figure}

In order to detect flare spikes automatically, we use our automated routine \textbf{which was} described in \cite{Goodarzi2019}. In this routine, first we extract gradual component ($F_{G}$) that is the main fluctuation component of the light curve due to the rotation modulation by the magnetic features, see middle panels of figure \ref{Fig1}. Then, by subtracting normalized gradual component ($f_{G}$) from the normalized flux (f), we obtain flare component ($f_{F}$), see bottom panels of figure \ref{Fig1}. We select the three times of the standard deviation of running difference ($3*\sigma$) as the threshold value for flare detection (red horizontal lines in the bottom panels of figure\ref{Fig1}). For more details about the procedure, please see paper I.
Eventually, a total number of 86067 flares are found on the 352 subgiant stars.  

\section{Magnetic features and flare parameters} \label{sec:param} 
We follow the definition of the magnetic parameters and flare indexes in paper I. Here, for ease of reference, we review them briefly.
From the light curve modulation, it is possible to evaluate stability of the magnetic features; long lived magnetic structures cause stable variation shape of the light curve. Autocorrelation index $i_{\rm AC}$ estimates this similarity on the light curve, which is defined
for an observed time series with N data points $\{X_t, t=0, 1, \ldots, N-1\}$ as:
\begin{equation}\label{equ:acf}
\rho(h)=\frac{\sum_{t=0}^{N-1-h} (X_{t+h}-\overline{X})(X_t-\overline{X})}{\sum_{t=0}^{N-1} (X_t-\overline{X})^2},
\qquad 0\leqslant h \leqslant N-1,
\end{equation}
\begin{equation}\label{equ:iac}
i_{\rm AC}=\frac{2}{N} \int_{0}^{N/2} |\rho(h)| dh,
\end{equation}

Where $\rho(h)$ is auto-correlation coefficient at the time lag h, $\overline{X}$ is the mean value of $X_t$, and N/2 in the integration limits is the largest possible value that permits the entire light curve to be included in the calculations. As in paper I, only the stars with rotation period shorter than 45 days are selected, because the length of each quarter is 90 days and the light curve of longer period stars could not be cross-correlated with itself and $i_{\rm AC}$ does not yield the correct value \citep{He2018} . 

The next magnetic parameter is effective range of light-curve variation ($R_{eff}$) which reflects the relative coverage and contrast of the magnetic structures on the stellar surface. For $X_{t}$ time series, $R_{eff}$ is calculated as below:  
 
\begin{equation}\label{equ:x_t}.
x_t=\frac{X_t-\widetilde{X}}{\widetilde{X}},
\qquad t=0, 1, \ldots, N-1,
\end{equation}
\begin{equation} \label{equ:reff}
R_{\rm eff}=2\sqrt{2} \cdot x_{\rm rms} \cdot [{1-(\frac{T_{spot}}{T_{phot}})^4}] =2\sqrt{2} \cdot \sqrt{\frac{1}{N} \sum_{t=0}^{N-1} x_t^2} \cdot [{1-(\frac{T_{spot}}{T_{phot}})^4}]
\end{equation}

where $T_{phot}$ and $T_{spot}$ are the temperature of the photosphere and the spot, respectively. We estimate the spot temperature $T_{spot}$ from the figure 7 of \cite{Berdyugina2005} and we use $T_{eff}$ from the Kepler input catalog to approximate $T_{phot}$. First term in equation \ref{equ:reff} represents relative coverage of the magnetic features and second term evaluates their contrast or level of darkness which depends on the temperature difference between stellar spots and the surrounding photosphere. Please note that we compute $i_{\rm AC}$ and $R_{eff}$ from the normalized gradual component $f_{G}$.

Furthermore, we define three parameters to evaluate the flare characteristics quantitatively. The first one is time occupation ratio of flares ($R_{\rm flare}$) which is defined as the ratio of occupied time by flares ($t_{\rm flare}$) to the observation time ($t_{\rm obs}$);

\begin{equation} \label{equ:Rflare}
R_{\rm flare}=t_{\rm flare}/t_{\rm obs}
\end{equation}

$R_{\rm flare}$ is a measure of the flares occurrence frequency and takes higher value when the star produces flares more often.

In order to calculate the second flare index, total power ($P_{\rm flare}$), first we should estimate the total energy released  by all identified flares ($U_{\rm flare}$):

\begin{equation}\label{equ:Uflare}
U_{\rm flare} = \int_{\rm flare} L_{\rm flare}(t)~dt~
\end{equation}

where L is the bolometric flare luminosity,

\begin{equation} \label{equ:Lflare}
L_{\rm flare} = \sigma_{\rm SB} T_{\rm flare}^4 A_{\rm flare}~
\end{equation}

In equation \ref{equ:Lflare}, $\sigma_{SB}$ is the Stefan–Boltzmann constant, $A_{\rm flare}$ is the flare area and $ T_{flare} $ is the flare temperature, which we assume to be 9000 K as its black body temperature \citep{HawleyFisher92,Kretzschmar2011}. Finally the total power of flares can be calculated as:

 \begin{equation} \label{equ:Pflare}
 P_{\rm flare}=U_{\rm flare}/t_{\rm obs}
 \end{equation}

The last flare index is the averaged flux magnitude of flares ($M_{\rm flare}$), that is the ratio of the total energy produced by all of flares to the flare time, 

\begin{equation} \label{equ:Mflare}
M_{\rm flare}=\frac{1}{t_{\rm flare}} \int L_{\rm flare}(t) dt=U_{\rm flare}/t_{\rm flare}
\end{equation}

To better understand the mentioned magnetic proxies and flare indexes, we have shown the light curves of two Kepler targets in figure\ref{Fig1} and their value of parameters in the first two rows of table\ref*{tab:parameters}. Panel (a) of figure \ref{Fig1} shows the light curve of KIC 6292732 which is a less active star in comparison with KIC 11957046 depicted in panel (b). 
The range of light curve variation in panel (a) of figure \ref{Fig1} is much lower than panel (b) and as a result, the value of $R_{eff}$ for KIC 6292732 is approximately 10 times less than KIC 11957046. Furthermore, the number and height of the flare spikes in panel (a) is less than panel (b), so the flare frequency, power and magnitude of KIC 6292732 is lower than KIC 11957046.

\floattable
\begin{deluxetable}{lccccccccccccccc}
	\tablecaption{Prarameters of Kepler subgiant flare stars. This table is available in its entirety (352 stars) and machine-readable form in the online journal.
		\label{tab:parameters}}
	\tablehead{
		\colhead{KIC} 
		&&  \colhead{$T_{\rm eff}$\tablenotemark{1}}  & \colhead{$P_{\rm rot}$\tablenotemark{2}} & 
		\colhead{$i_{\rm AC}$}  & \colhead{$R_{\rm eff}$} & \colhead{$R_{\rm flare}$}
		&  \colhead{$P_{\rm flare}$}  & \colhead{$M_{\rm flare}$} & \colhead{Thr\tablenotemark{3}} & \colhead{$N_{\rm f}$\tablenotemark{4}} \\
		&&                   (k) &                        (day)&              & $(10^{-2})$&
		$(10^{-2})$& $(10^{22})$        &$(10^{24})$  &$(10^{-2})$                         
	}
	\startdata
	6292732\tablenotemark{*} && 5861 & 11.25 & 0.350 & 0.503 & 0.307 & 0.162 & 0.615  & 0.057 & 163  && \\
	11957046\tablenotemark{**}&& 5855 & 5.95  & 0.359 & 5.013 & 2.206 & 3.287 & 1.733  & 0.191 & 465  && \\
	1995351 && 5627 & 3.268 & 0.232 & 1.119 & 0.638 & 0.141 & 0.238  & 0.062 & 233  && \\
	2018631 && 6221 & 4.268 & 0.268 & 0.018 & 0.187 & 0.070 & 0.364  & 0.027 & 102  && \\ 
	2158047 && 5288 & 5.424 & 0.261 & 1.122 & 0.765 & 0.207 & 0.283  & 0.086 & 285  && \\
	2158750 && 6131 & 7.777 & 0.264 & 0.091 & 0.498 & 0.324 & 0.722  & 0.056 & 217  && \\	
	2168263 && 6276 & 9.029 & 0.369 & 0.445 & 0.416 & 0.473 & 1.187  & 0.129 & 172  && \\
	2308980 && 5757 & 15.712& 0.269 & 0.203 & 0.215 & 0.111 & 0.519  & 0.085 & 106  && \\
	2446436 && 6063 & 0.803 & 0.282 & 0.719 & 0.349 & 0.333 & 1.121  & 0.068 & 145  && \\
	2712904 && 6231 & 2.522 & 0.266 & 0.892 & 0.270 & 0.293 & 1.172  & 0.139 & 118  && \\
	2719755 && 6058 & 3.895 & 0.269 & 1.116 & 0.783 & 0.669 & 1.298  & 0.061 & 312  && \\
	\enddata
	\tablenotetext{1}{From the Kepler Input Catalog \citep{Huber et al.14,Huber14}.}
	\tablenotetext{2}{From the rotation period catalog derived by \cite{McQuillan2014}.}
	\tablenotetext{3}{Threshold value of flare detection on the normalized flare component, see paper I.}
	\tablenotetext{4}{Number of flares.}
	\tablenotetext{*} {The light curve of this target is shown in panel (a) of Fig.\ref{Fig1}.}
	\tablenotetext{**} { The light curve of this target is shown in panel (b) of Fig.\ref{Fig1}.}
\end{deluxetable}

\section{Results}\label{sec:results} 

Magnetic parameters and flare indexes are calculated from the Q2-Q16 Kepler light curves for the 352 subgiants with spectral types F, G and K and their mean values over the 15 quarters are given in table \ref{tab:parameters}.
Since there are no published calculation of the Rossby number or convective turnover time for the SGB, like the one computed by \cite{wright2011} for the MS, we replace the Rossby number with the quantity ${P^{-2}R^{-4}}$ which was proposed by \cite{Reiners2014}. This quantity produces the minimum scatter of X-ray emission plots and can be an optimal alternative to the Rossby number. Furthermore, the role of rotation rate in driving dynamo is more important and more proven than the size or radius of the stars \citep{Reiners2014}. Owing to this fact, in our analysis we emphasize mostly on rotation period rather than the radius.
	  
\begin{figure}
	\epsscale{1.3}
	\plotone{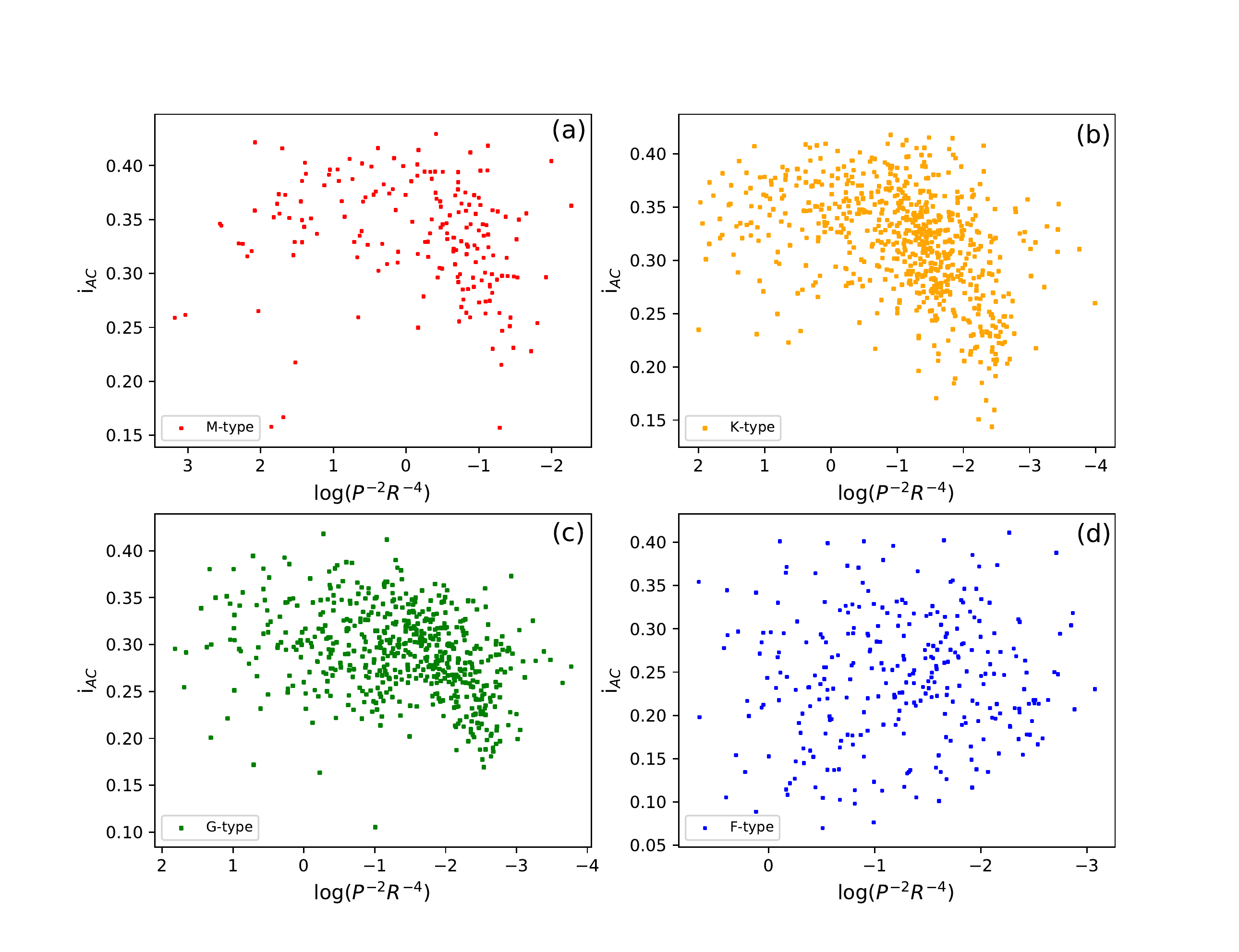}
	\caption{Auto correlation index ($i_{AC}$) versus $P^{-2}R^{-4}$ for the main sequence stars (reproduction of figure 2 in paper I). The four panels correspond to the four spectral types of stars: M-type stars (top left, red), K-type stars (top right, yellow), G-type stars (bottom left, green) and F-type stars (bottom right, blue).}
	\label{Fig2}
\end{figure}

Our goal is to compare the results of our analysis for the SGB with the MS in paper I. However, in paper I magnetic activity indicators were plotted versus Rossby number. Hence, we reproduce one of its figures versus ${P^{-2}R^{-4}}$ as an example to show that there is no big difference and we can do the comparison with confidence. Figure \ref {Fig2} is the reproduction of figure 2 in paper I, which displays autocorrelation index $i_{\rm AC}$ as a function of ${P^{-2}R^{-4}}$. The general behavior of both figures is similar and the very small details are negligible. 

Since less massive and cooler K and M dwarfs spend more time on the main-sequence \citep{Schwarzschild58,Kippenhahn&Weigert90,Hurley2000}, most of them have not reached to the SGB. So, in spite of the MS sample in the paper I which contains four spectral types (F, G, K and M), our sample of subgiants consists mostly of more massive G and F type with few K-type stars. Therefore, statistical conclusion is impossible for K-type stars. But, we keep them for qualitative interpretation. Also the spectral classification has been done using the same temperature criteria described in paper I. 

We estimate the error of the computed parameters using the formulas in the appendix of \cite{He2018} and apply the error propagation formula for additional terms. We show only the error bars of a small number of points as a representative in order to avoid cluttering the figures.     

\floattable
\begin{splitdeluxetable*}{lcccccccccccccccccccccBccc}
	\tablecaption{Magnetic parameters and flare indexes for the MS and SGB in each spectral type (break)\label{tab:parameters2}}
	\tablehead{
		\colhead{}  &&  \multicolumn{3}{c}{$i_{\rm AC}$}  &&  \multicolumn{3}{c}{$R_{\rm eff}$}  &&  \multicolumn{3}{c}{$R_{\rm flare}$} &&  \multicolumn{3}{c}{$P_{\rm flare}$} &&  \multicolumn{3}{c}{$M_{\rm flare}$} && \multicolumn{3}{c}{period (days)} \\
		&&                         &                        &              
		&&                         &  $(10^{-2})$           &             
		&&                         &      $(10^{-2})$       & 
		&&                         &     $(10^{22})$        & 
		&&                         &     $(10^{24})$        &               &\\ 
		\cline{3-5} \cline{7-9} \cline{11-13} \cline{15-17} \cline{19-21} \cline{23-25}
		&&       \colhead{F}       & \colhead{G}           &    \colhead{K}
		&&       \colhead{F}       & \colhead{G}           &    \colhead{K}
		&&       \colhead{F}       & \colhead{G}           &    \colhead{K} 
		&&       \colhead{F}       & \colhead{G}           &    \colhead{K} 
		&&       \colhead{F}       & \colhead{G}           &   \colhead{K}
		&&       \colhead{F}       & \colhead{G}           &   \colhead{K}               
	}
	\startdata
	Main Sequence\tablenotemark{1}  &&  0.243 & 0.290 &  0.315  && 0.63  & 1.33 &  1.43  && 0.40  & 0.60 & 0.76 && 0.258 & 0.604 & 0.606 && 0.801 & 0.996 & 0.861 && 4.46& 9.03&13.39\\
	Sub-Giant && 0.273  & 0.310 & 0.345 && 0.75  & 2.06 & 3.74 && 0.47 & 0.90 & 1.24 &&  0.671 & 1.435 & 2.082 && 1.178 & 1.532 & 2.241 &&4.95 & 6.22&10.34\\  
	\enddata
	\tablenotetext{1}{The average value of parameters for the Main sequence stars is computed from the table.1 of \cite{Goodarzi2019}.}
\end{splitdeluxetable*}

\subsection{Comparison with the main-sequence stars} \label{subsec:compare}

For general comparison between the magnetic activity of the MS stars which was reported in paper I and the SGB stars that are discussed in this paper, the average values of the magnetic parameters and flare indexes for these two evolutionary states are given in table \ref{tab:parameters2}. In order to avoid the effect of different sample selection in the MS and SGB (lack of M-type stars and low number of K-type stars in the SGB sample), we present this table with spectral type classification.
 
From table \ref{tab:parameters2} it can be seen that the autocorrelation index of the SGB is approximately the same as the MS. However, effective range of light curve fluctuations increases considerably; $R_{eff}$ is higher in the SGB than the MS for the three spectral types. The increase of the relative coverage and contrast of the magnetic features in the SGB is due to the deepening of the convective zone \citep{Van Saders2019} which increases the production of magnetic field \citep{Jenkins2011}. From production of more magnetic field, we expect more stable magnetic structures \citep{Hall94}. However, autocorrelation index of the SGB is not influenced by this fact, because more vigorous convective turbulence probably cause to decay of star spots and annihilate the effect of increased lifetime resulted from their stronger magnetic field.  

It can also be seen from table \ref{tab:parameters2} that the time occupation ratio ($R_{flare}$), total power ($P_{flare}$) and averaged flux magnitude of flares ($M_{flare}$) in SGB are higher than in the MS. The average rotation period of MS sample in paper I is 10.07 days and in the SGB sample is 6.06 days. The reason for the lower average rotation period in the SGB is the existence of hot earlier type stars with faster rotation in this sample \citep{Garcia2014}. In other words, the increase of the stellar moment of inertia in the SGB results in only moderate spin-down of the stars \citep{Gilliland85}. On the other hand, SGB stars have deeper convective envelopes in comparison with the MS stars \citep{Van Saders2013}. So, the Rossby number $Ro=P_{rot}$/$\tau_{conv}$ (where $P_{rot}$ is the rotation period and $\tau_{conv}$ is the convective turnover time) decreases and cause to more vigorous magnetic field production. This causes to higher values of accumulated magnetic energy on one hand and the increase of the possibility of re-connection on the other hand. Therefore, we see grater values of magnetic activity indicators such as relative coverage and contrast of the magnetic structures and also frequency, total power and magnitude of flares.

\cite{Simon&Drake89} analyzed ultraviolet observations of the late-type subgiants and concluded that the activity of low-mass stars with M $<$ 1.25 \(M_\odot\) do not show any large increase or decrease once they leave the main sequence; however in stars with M $>$ 1.25 \(M_\odot\), the UV emission declines when the star evolves to the SGB. This is inconsistent with our results in which the magnetic activity indicators increase for our SGB sample with mass range of 0.8 \(M_\odot\) $<$M $<$1.25 \(M_\odot\). Furthermore, using the stellar evolution codes and the Ca II emission lines associated with the chromospheric activity, \cite{Gilliland85} concluded that in the evolution path of the stars that are slightly more massive than the sun, the increase of the convective turnover time and the subsequent decrease of Rossby number is more important than the increase of moment of inertia, thus magnetic activity enhances strongly with age.     

\subsection{Magnetic features stability, coverage and contrast in subgiants} \label{subsec:Magnetic} 

\begin{figure}
	\gridline{\fig{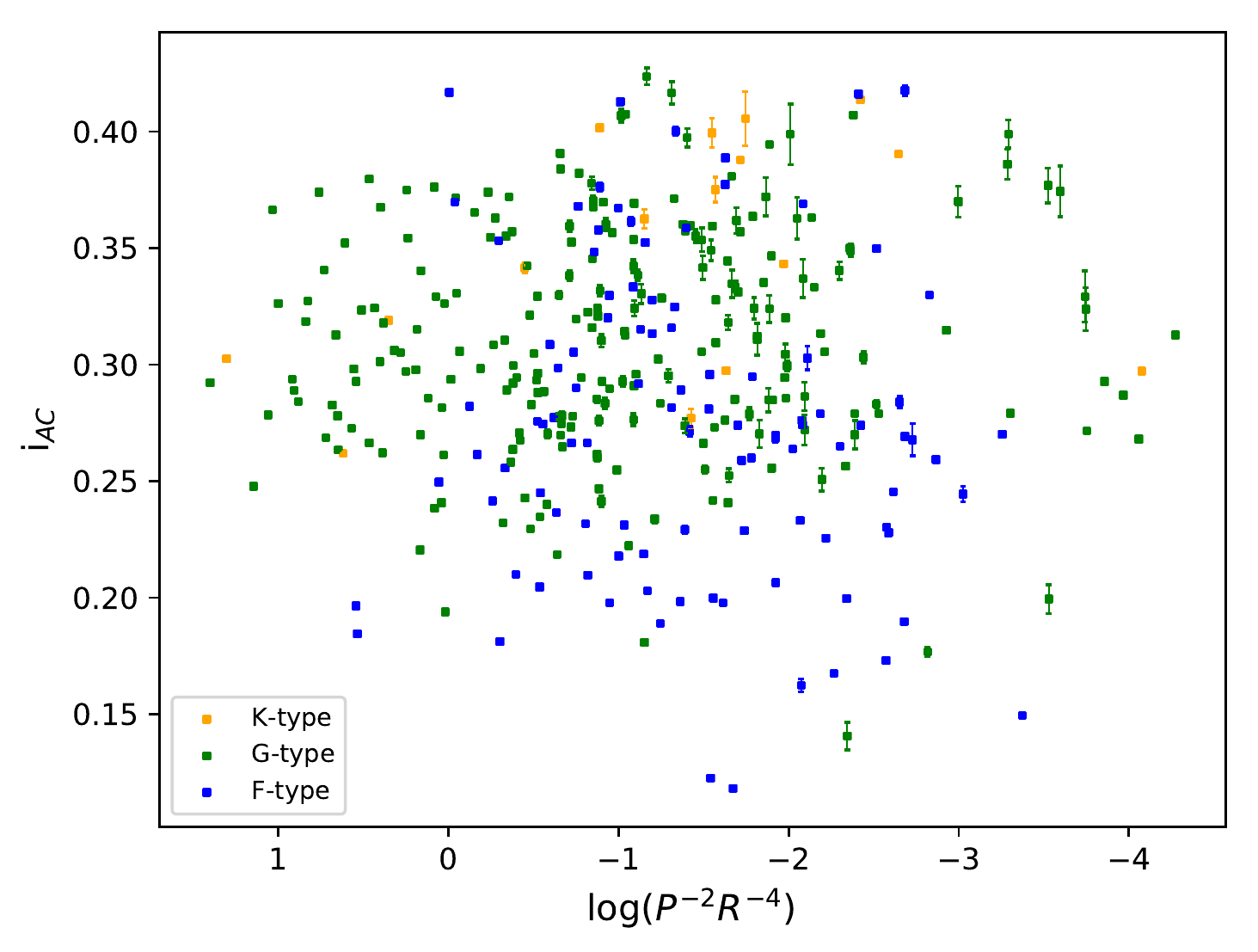}{0.51\textwidth}{(a)}
		\fig{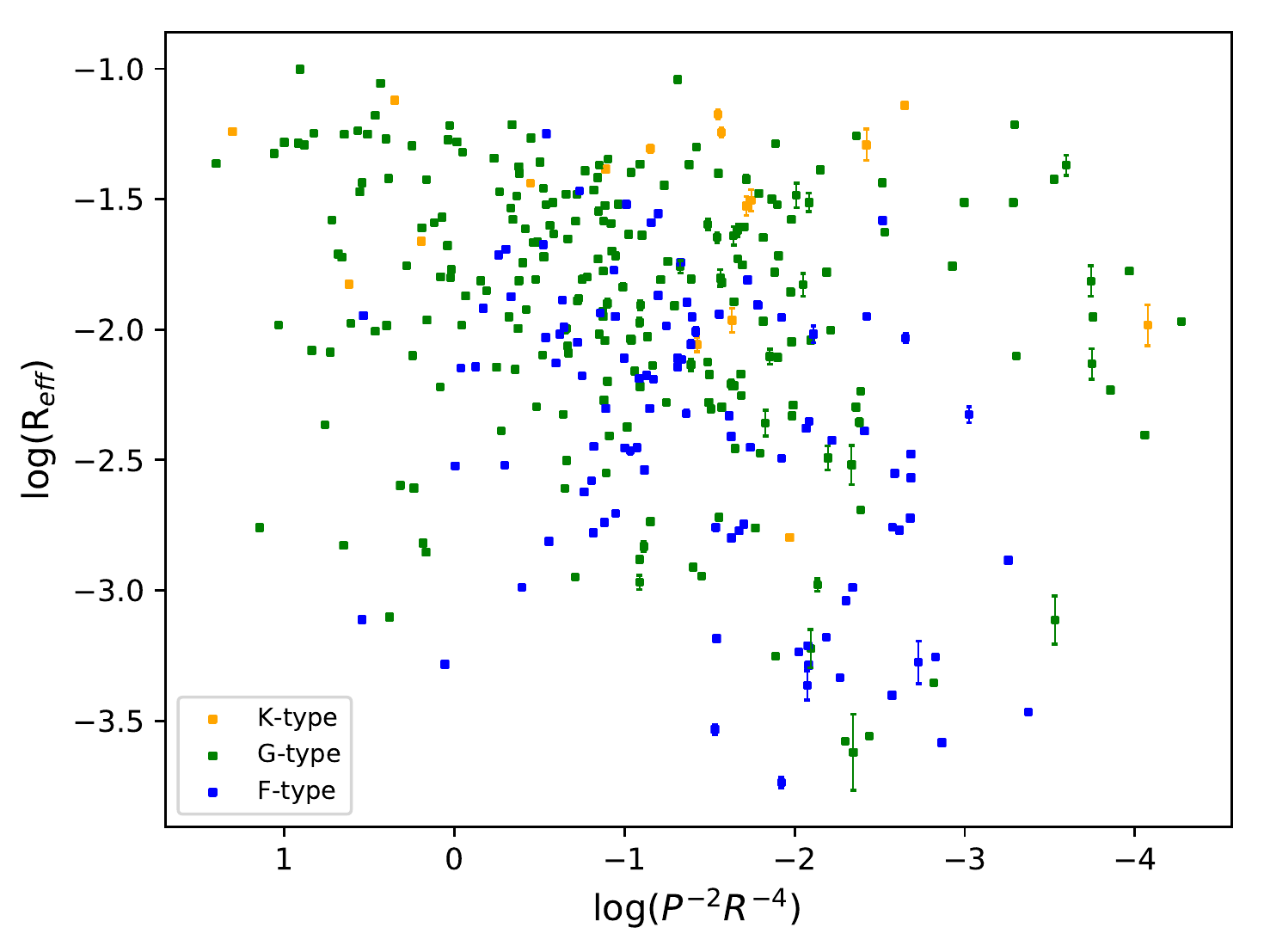}{0.51\textwidth}{(b)}}
	\caption{Autocorrelation index $i_{AC}$ (panel a, left) and effective range of light curve variation $R_{eff}$ (panel b, right) as a function of ${P^{-2}R^{-4}}$ in logarithmic scale for the SGB. Colors indicate spectral types of the stars.}
	\label{Fig3}
\end{figure}

Panel (a) and (b) of figure \ref{Fig3} displays autocorrelation index $i_{AC}$ and effective range of light curve variations $R_{eff}$ as a function of ${P^{-2}R^{-4}}$ in logarithmic scale for three spectral types. As discussed before, $i_{AC}$ reflects magnetic feature stability and $R_{eff}$ evaluate relative coverage of the magnetic structures on the stellar surface together with their contrast.
We show the horizontal axis inversely so that, as in the paper I, the movement in the right direction corresponds to an increase in the rotation period of the stars.

\begin{figure}
	\epsscale{0.8}
	\plotone{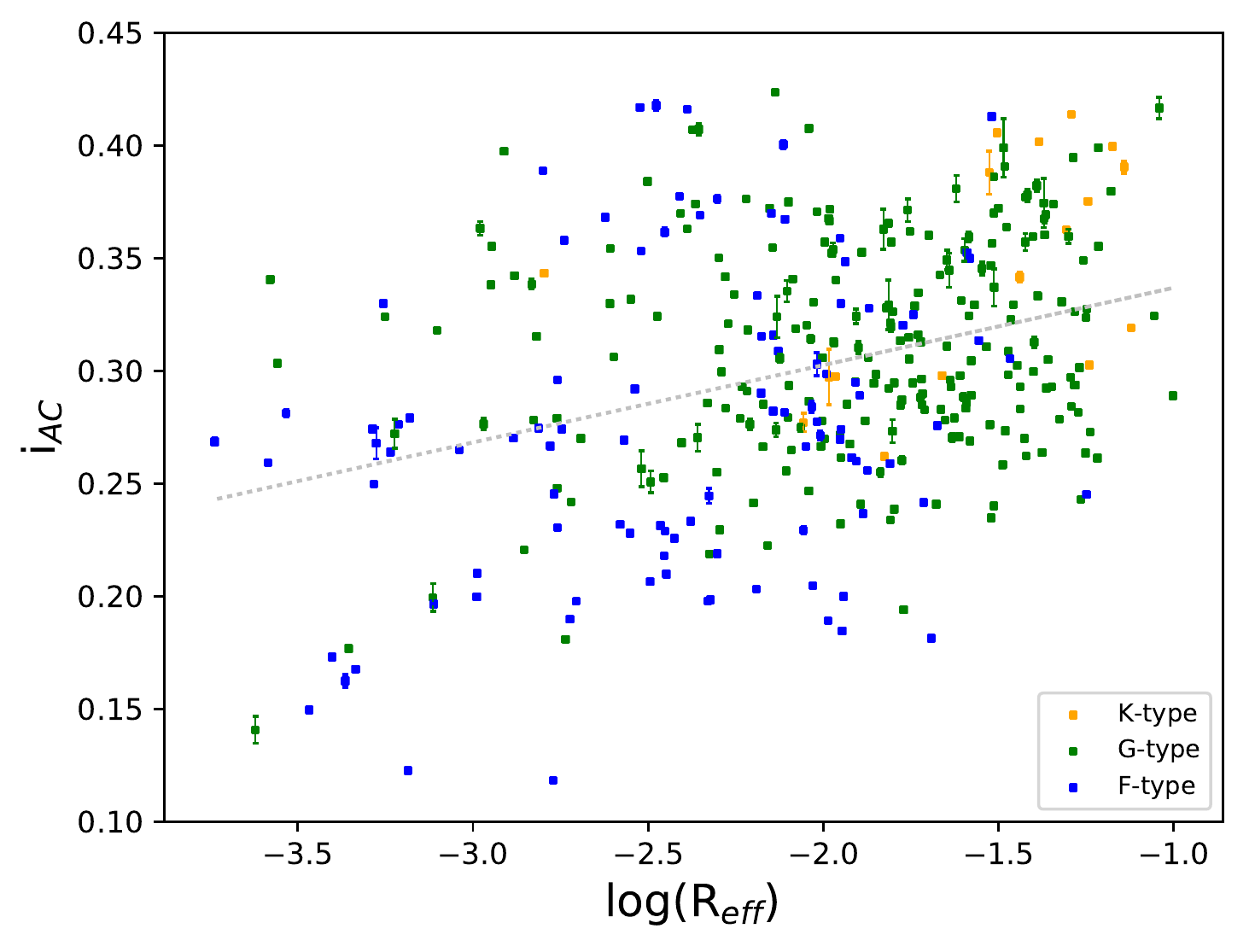}
	\caption{Auto correlation index ($i_{AC}$) versus effective range of light curve variations ($R_{eff}$). Colors indicate spectral type of the stars.}
	\label{Fig4}
\end{figure}

Both panels of figure \ref{Fig3} indicate that there is no obvious dependency to ${P^{-2}R^{-4}}$ quantity in the SGB; neither $i_{AC}$ nor $R_{eff}$ has a dominant increase or decrease. This is the opposite of what we observed in the MS sample (see figure 2 and 3 of the paper I) in which the magnetic feature stability, coverage and contrast of G-, K-, M-type stars increase with decreasing Rossby number until reaching to saturation level; however F-type stars do not show such dependency. Here, for the subgiants, even G-type stars magnetic proxies do not show saturation behavior. It means that, in spite of what we expect from dynamo mechanism, magnetic feature stability, coverage and contrast are an independent parameters of rotation rate in the SGB. So, in this evolutionary state of the stars, faster rotation does not necessarily produce more stable and stronger magnetic fields.

In figure \ref{Fig4} we plot the autocorrelation index as a function of the effective range of light curve variation. Stability of the magnetic features increase by increasing their relative coverage and contrast, however the slope of this increment is less than the MS stars; gray dashed line shows linear least squares fit for SGB stars, which has smaller slope (0.034) in comparison with the main sequence stars (0.058). As vigor of convection is higher in the SGB than in the MS, magnetic feature stability is probably affected by this fact and less influenced by the size and contrast of the magnetic structures.   
    
\subsection{Flare Indexes of Subgiant Stars } \label{subsec:index}

\subsubsection{Time Occupation Ratio of Flares} \label{subsubsec:Rf}

\begin{figure}
	\epsscale{0.8}
	\plotone{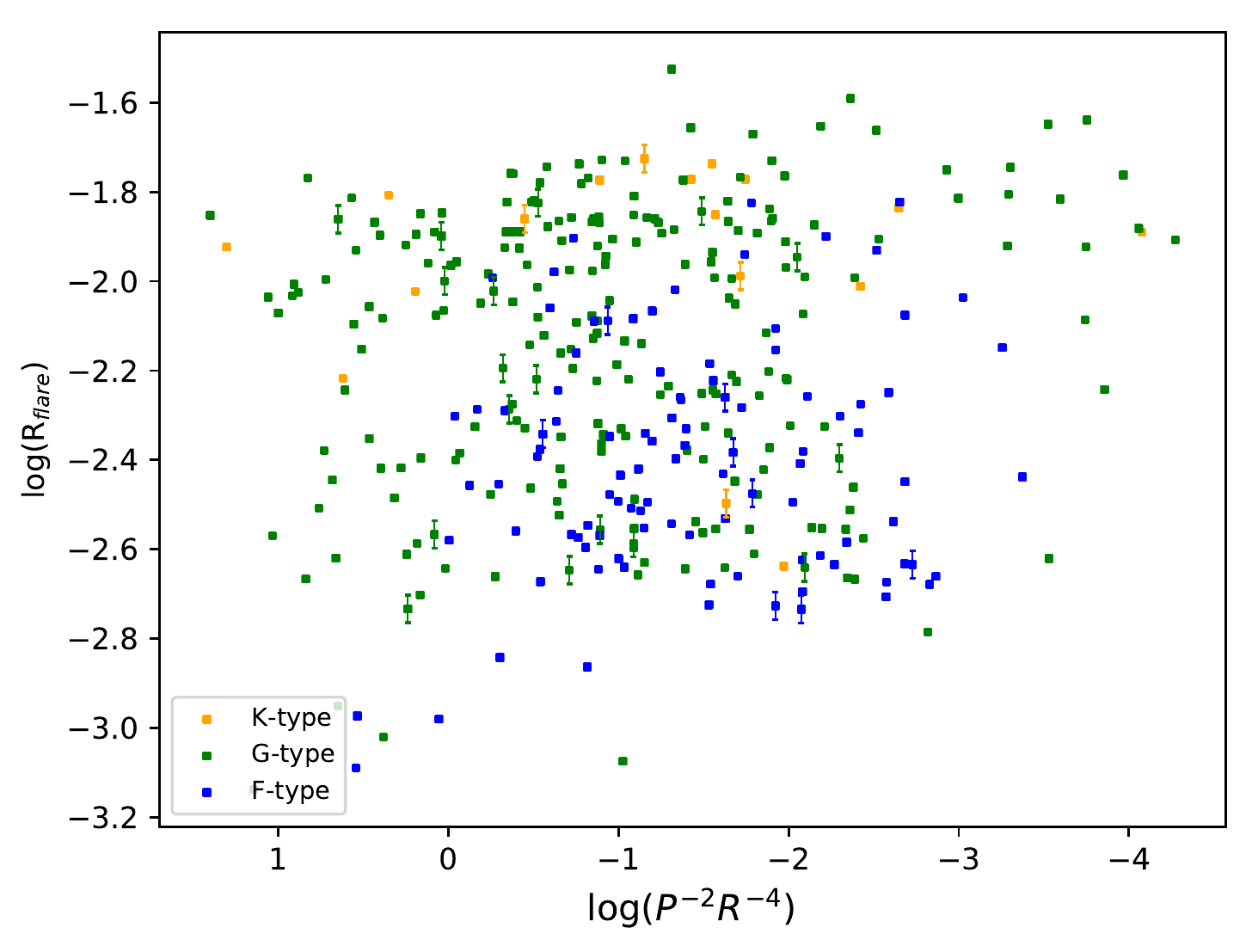}
	\caption{Time occupation ratio of flares ($R_{\rm flare}$) versus logarithm of ${P^{-2}R^{-4}}$ for different spectral types.}
	\label{Fig5}
\end{figure}

In figure \ref{Fig5} we plot the time occupation ratio of flares versus logarithm of ${P^{-2}R^{-4}}$ for three spectral types. It can be seen that unlike the MS (see figure 5 of paper I), $R_{\rm flare}$ does not have a dominant increase or decrease behavior in the SGB and instead is scattered everywhere. This means that the variation of the flare rate in the SGB is not affected very much by the rotation period and the size of the stars.

The level of chromospheric activity and its relation with the Rossby number was studied in a sample of 121 single stars from the SGB by \cite{duNascimento03}. They found that only in stars with $B-V<0.55$ (see their figure 7) there is no dependency of CaII emission flux to the Rossby number and in the rest (stars with $B-V>0.55$) chromospheric activity increases by decreasing Rossby number. However, it should be noted that they computed convective turnover time of the SGB stars from the MS relation in \cite{Noyes84} that is not very accurate due to the deepening of the convective envelope on the SGB. In our analysis, which most of the stars have $0.4<B-V<0.8$, none of the magnetic activity indicators show dependence on the ${P^{-2}R^{-4}}$ parameter (please see also sections \ref{subsubsec:Pf} and \ref{subsubsec:Mf}).

\begin{figure}
	\gridline{\fig{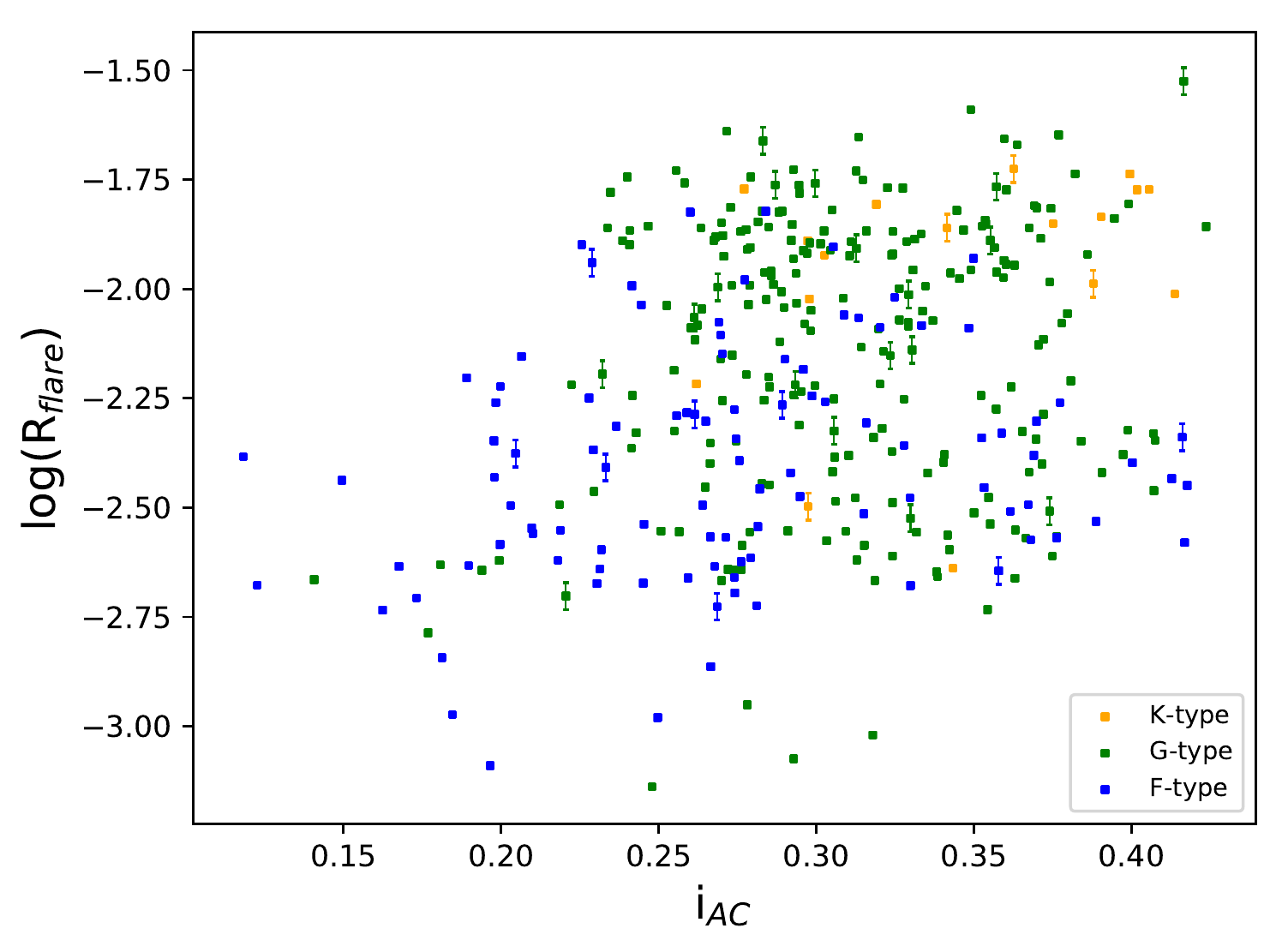}{0.5\textwidth}{(a)}
		\fig{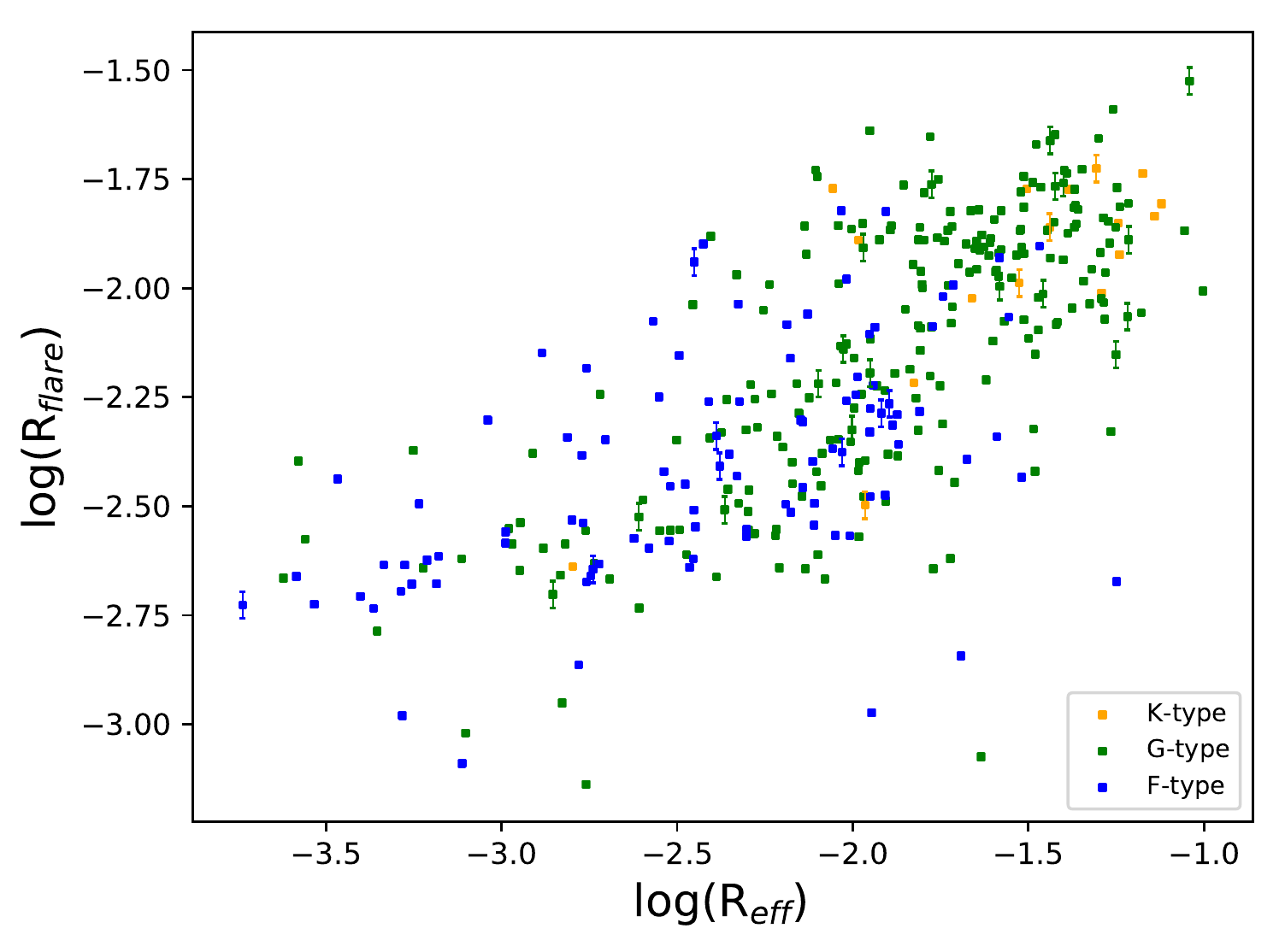}{0.5\textwidth}{(b)}}
	\caption{Logarithm of time occupation ratio of flare ($R_{\rm flare}$) versus autocorrelation index $i_{AC}$ (panel a, left) and versus effective range of light curve fluctuations $R_{eff}$ (panel b, right) for different spectral types that are indicated by colors.}
	\label{Fig6}
\end{figure}

Figure \ref{Fig6} displays $R_{\rm flare}$ versus autocorrelation index (panel a, left) and versus effective range of light curve fluctuations (panel b, right). Like the main sequence stars (see figure 6 of paper I), there is no strict increase of flare frequency with magnetic feature stability. However, the relative coverage and contrast of the magnetic structures plays more important role and $R_{\rm flare}$ increases clearly with increasing $R_{eff}$. When the coverage of magnetic features increases on the stellar surface, the possibility of reconnection and its consequent flare event increases. 

\subsubsection{Total power of flares} \label{subsubsec:Pf}
\begin{figure}
	\epsscale{0.8}
	\plotone{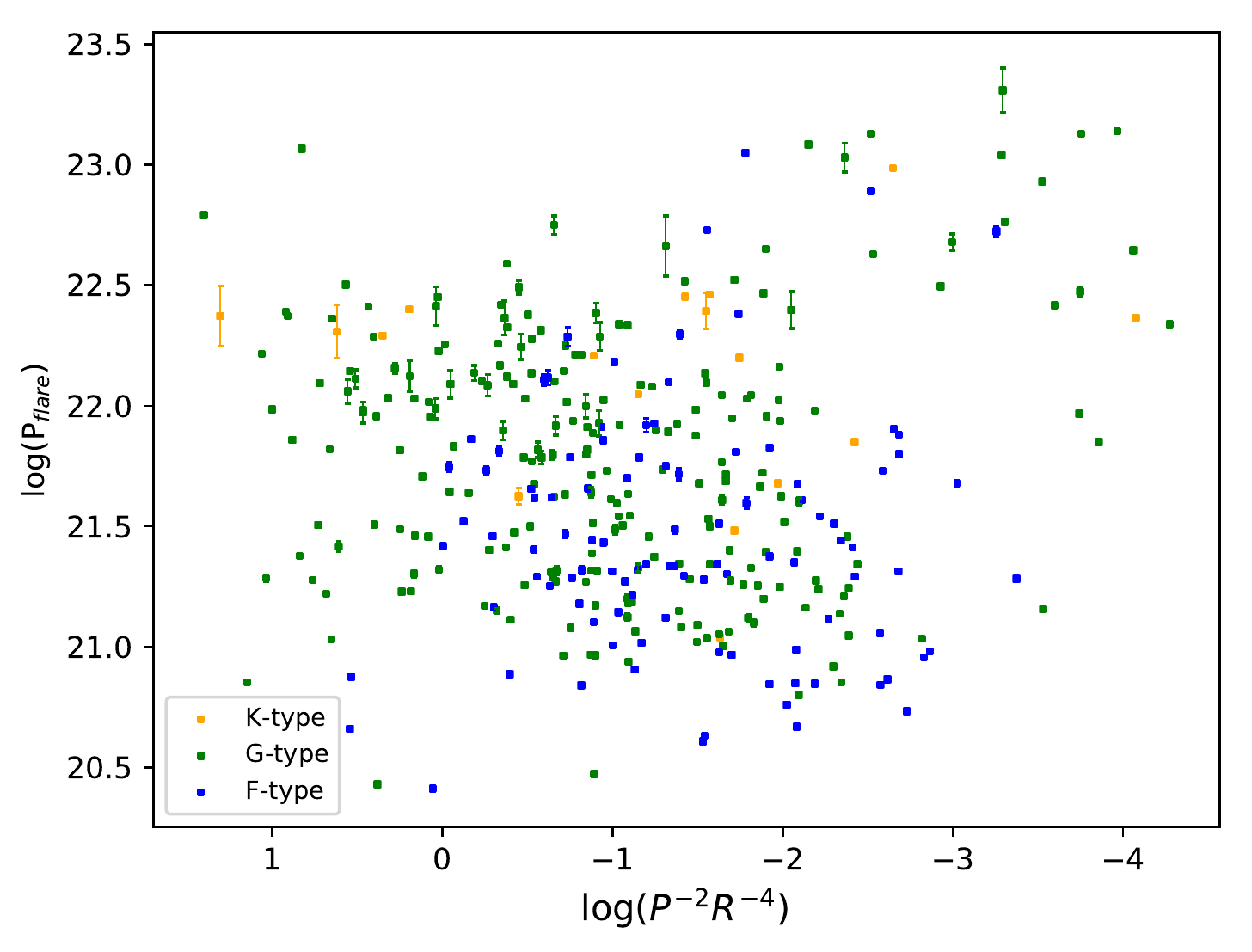}
	\caption{Total power of flares ($P_{flare}$) versus the logarithm of ${P^{-2}R^{-4}}$ for different spectral types.} 
	\label{Fig7}
\end{figure}

As defined in section \ref{sec:param}, $P_{\rm flare}$ gives an evaluation of the total power produced by all diagnosed flares. Figure \ref{Fig7} is $P_{\rm flare}$ versus logarithm of $P^{-2}R^{-4}$. Again, unlike the MS stars, no rotational velocity relationship could be recognized from these three spectral types. It can be concluded that dynamo procedure works with different mechanisms when the stars evolve to the SGB, in which there is not strict rotation rate dependency and saturation regime; possibly due to the deepening of the convective zone and magnetic field production, which is already in its maximum level and does not increase further when rotation rate increases. 

\begin{figure}
	\gridline{\fig{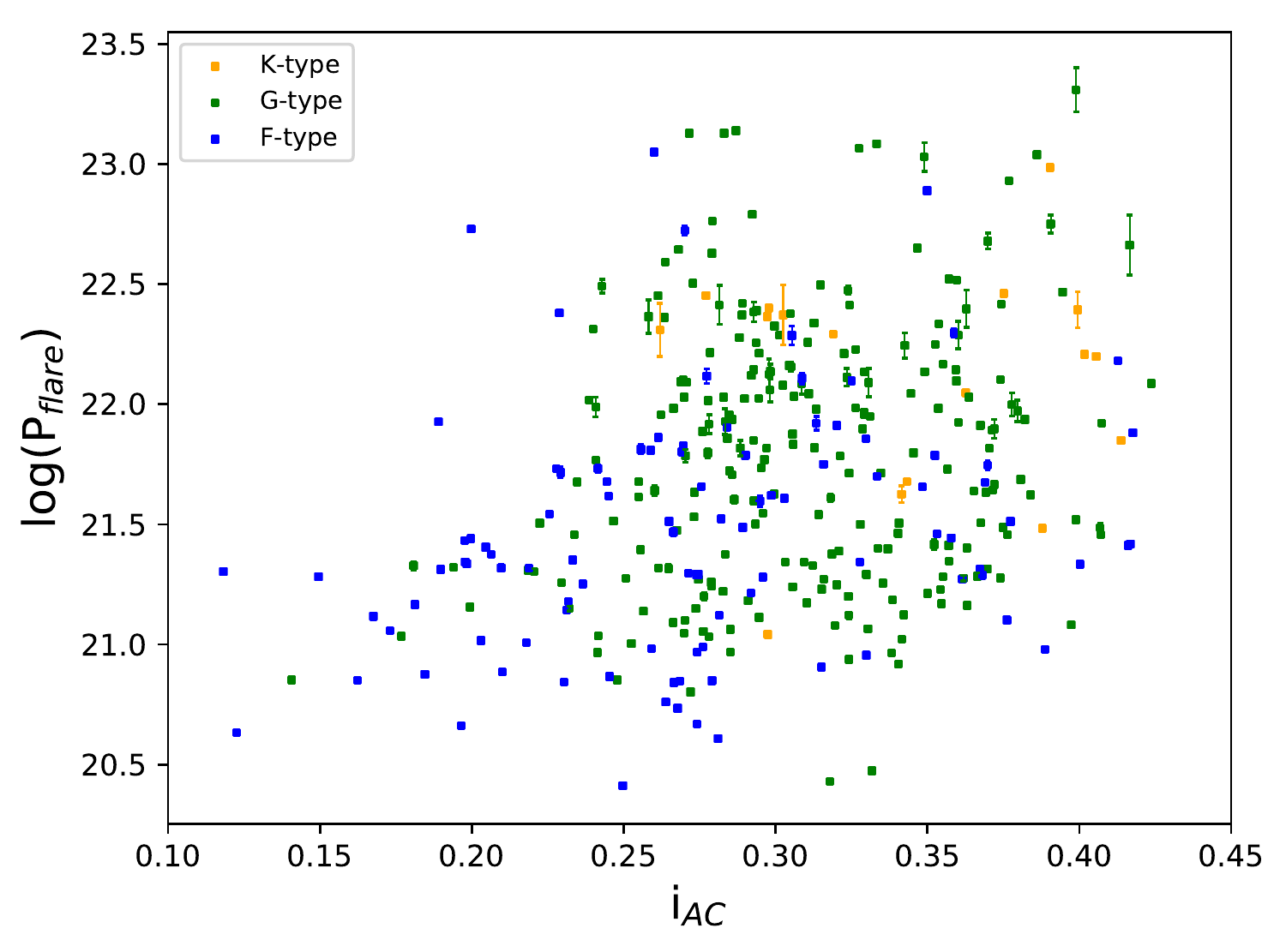}{0.5\textwidth}{(a)}
		\fig{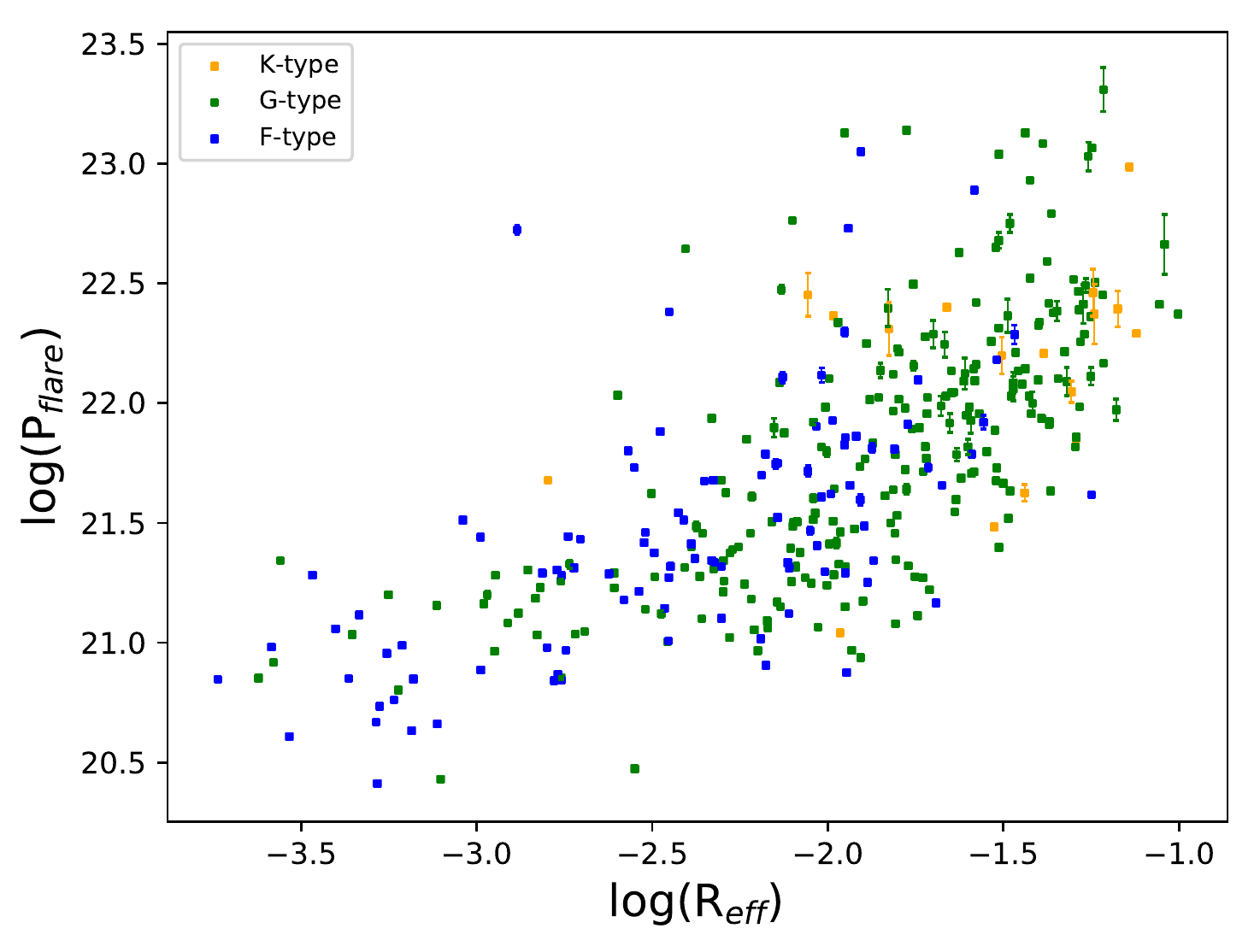}{0.5\textwidth}{(b)}}
	\caption{Total power of flares ($P_{flare}$) vs. the autocorrelation index $i_{AC}$ (panel a, left) and vs. the effective range of light-curve fluctuations $R_{eff}$ (panel b, right) for subgiant stars from different spectral types.}
	\label{Fig8}
\end{figure}

Panel (a) and (b) of figure \ref{Fig8} show the total power of flares versus $i_{AC}$ and $R_{eff}$, respectively. Panel (a) is more scattered than having trends, but increase of $P_{\rm flare}$ with increase of $R_{eff}$ in panel (b) is quite noticeable for all of the three spectral types. Since the magnetic energy is proportional \textbf{to} the strength and volume of magnetic structures, larger and darker stellar spots can store more magnetic energy inside themselves \citep{Notsu2013,Notsu2019} and release it in flare events. This is similar to what we saw for the MS stars in paper I.

\subsubsection{Averaged Flux Magnitude of Flares} \label{subsubsec:Mf}
\begin{figure}
	\epsscale{1.3}
	\plotone{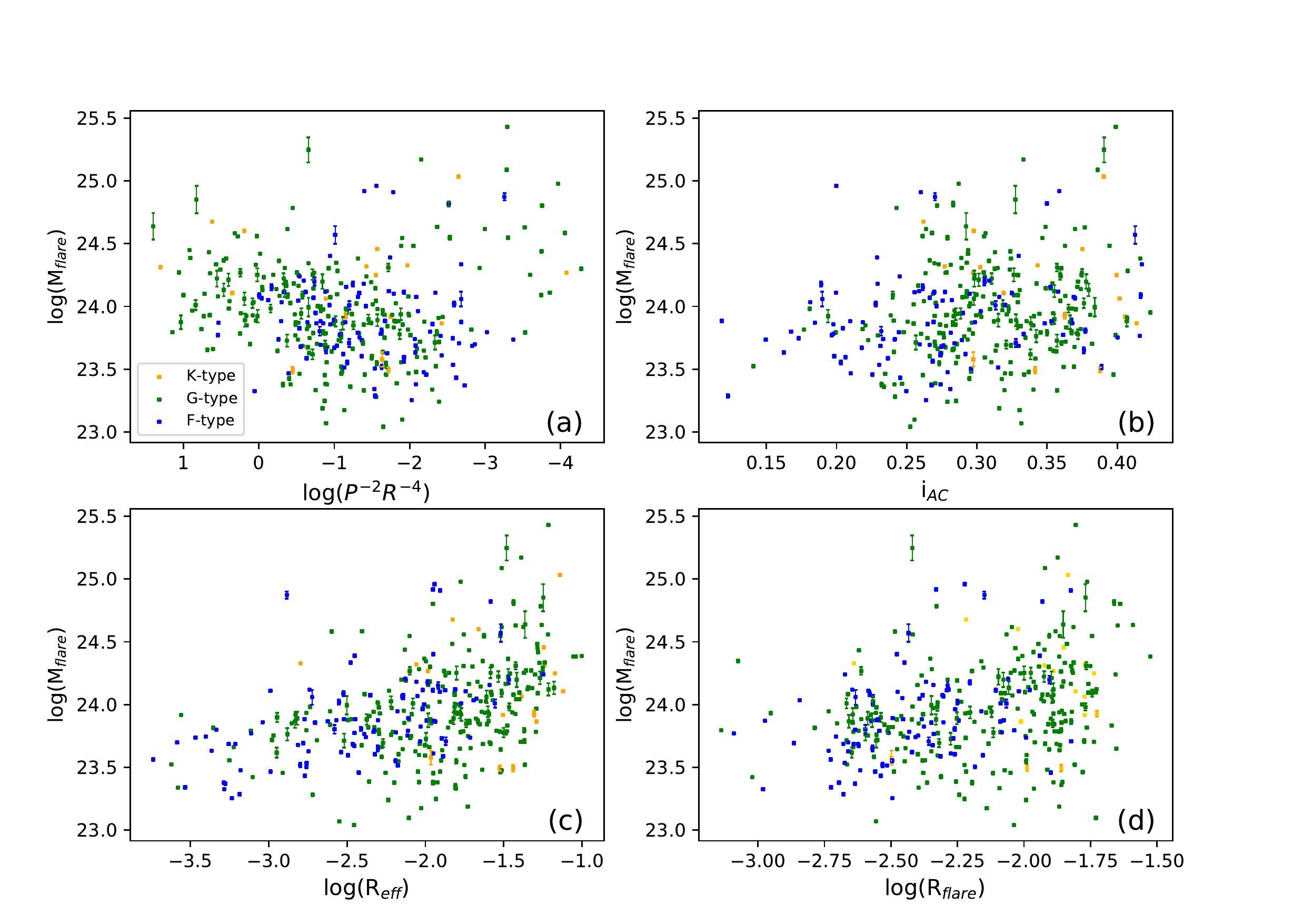}
	\caption{Averaged flux magnitude of flares ($M_{\rm flare}$) as a function of: ${P^{-2}R^{-4}}$ (panel a), autocorrelation index $i_{AC}$ (panel b), effective range of light curve fluctuations $R_{eff}$ (panel c) and time occupation ratio of flares $R_{\rm flare}$ (panel d) for different spectral types which are indicated by colors.}
	\label{Fig9}
\end{figure}

Figure \ref{Fig9} provides averaged flux magnitude of flares as a function of different variables. Panel (a) shows $M_{\rm flare}$ versus the logarithm of $P^{-2}R^{-4}$; there is a faint increase of $M_{\rm flare}$ towards the left part of the panel which corresponds to the shorter rotation period, i.e., stars that have faster rotation can produce flares with higher magnitude. This is the same as what we observed for the MS in paper I.  

Panel (b) of figure \ref{Fig9} displays $M_{\rm flare}$ as a function of autocorrelation index. Averaged flux magnitude of flares at higher value of $i_{AC}$ increases slightly, because more durable magnetic structures are likely to store more magnetic energy within themselves.  
 
In panel (c) of figure \ref{Fig9} we see averaged flux magnitude of flares versus effective range of light curve fluctuations which increase clearly with increasing $R_{eff}$. Especially beyond $log(R_{eff}$)= -2.0, the slope of scatter plot increases suddenly, and panel (b) of figures \ref{Fig6} and \ref{Fig8} show the same behavior. This threshold value of $R_{eff}$ has already been seen in the MS stars (see figures 6, 8, and 9 of paper I), and implies that when the relative coverage and contrast of the magnetic features exceeds a certain limit, the frequency, power and magnitude of flares \textbf{increase} suddenly. 
        
Finally, panel (d) of figure \ref{Fig9} is $M_{\rm flare}$ versus time occupation ratio of flares. There is a weak trend of enhancement of flare magnitude with increasing of $R_{\rm flare}$ suggesting that large flares are more likely to occur in stars which produce flare more often. 
  
Generally, the behavior of $M_{\rm flare}$ versus different variables for the SGB stars is the same as for the MS stars. The only difference is the lack of the M-type stars and small number of K-type stars; both of which have low values of flux magnitude. In other words, the maximum flare magnitude on the SGB is the same as the MS; but the minimum value is higher than the MS due to the lack of late type stars.

\subsection{The Amount of Convective Matter} \label{subsubsec:massconv} 
It is suggested that in addition to the role of the rotation rate in production of magnetic field, the amount of the convective matter may have important effect. \cite{Pettersen89} considered this for the MS and concluded that there is a good correspondence between flare luminosity of red dwarfs and the volume of the convective zone. However, they did not found such a positive relationship with the mass of the convective zone.

\begin{figure}
	\epsscale{0.7}
	\plotone{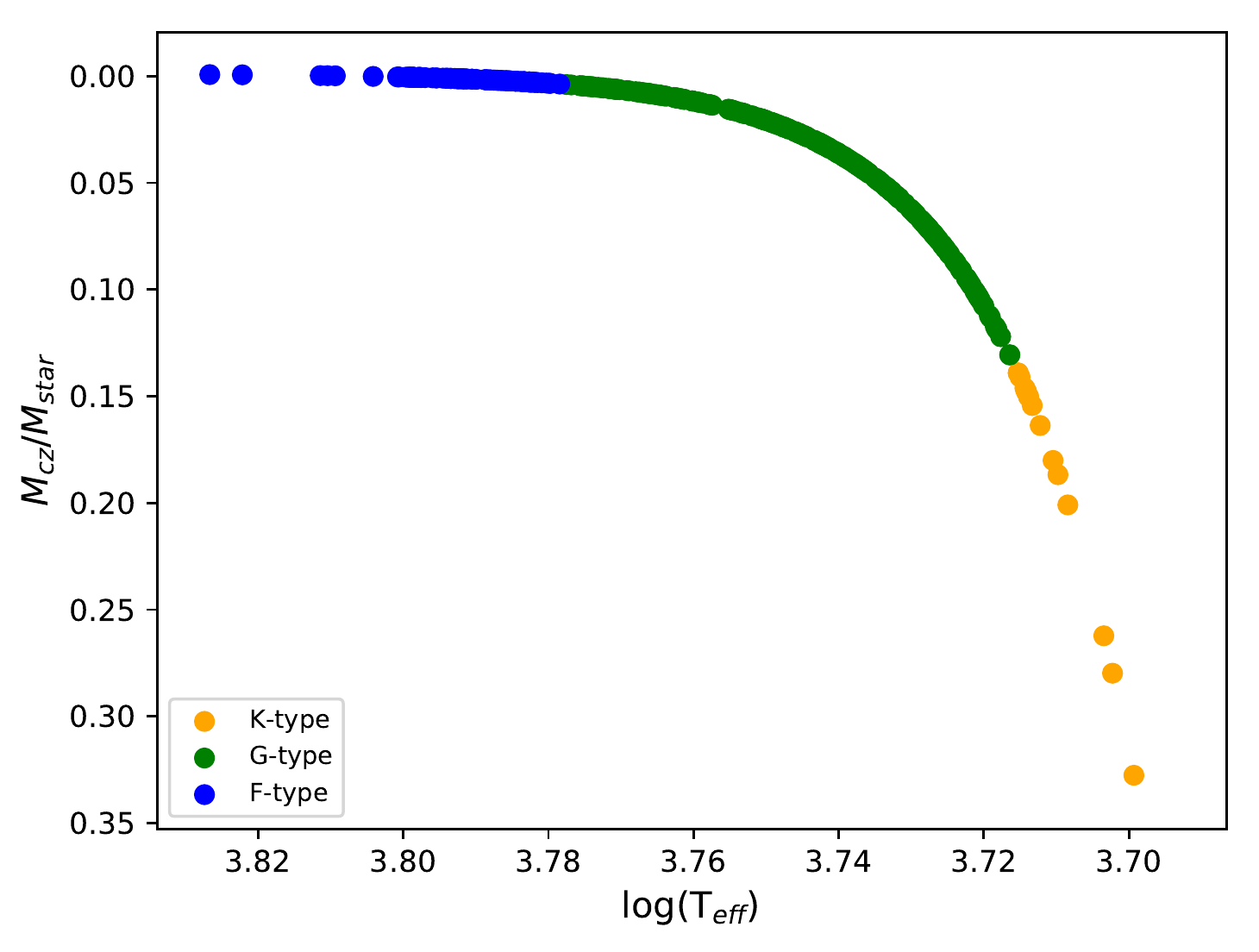}
	\caption{The relative mass of the convective zone ($M_{\rm CZ}/M_{\rm star}$) versus logarithm of effective temperature for the SGB sample.}
	\label{Fig10}
\end{figure}

\begin{figure}
	\epsscale{1.3}
	\plotone{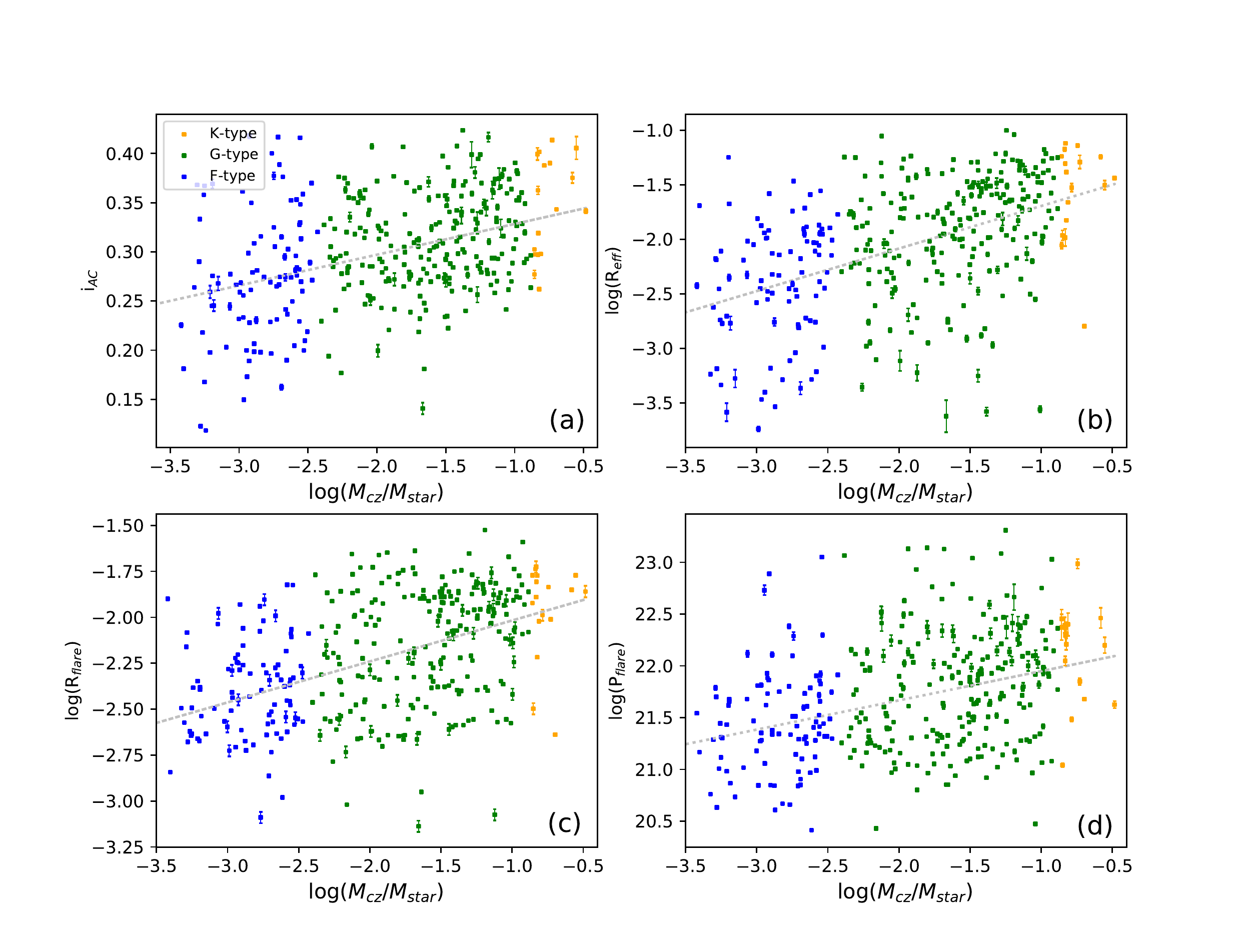}
	\caption{Variation of: (a) autocorrelation index $i_{AC}$ , (b) effective range of light curve fluctuations $R_{eff}$ , (c) time occupation ratio of flares $R_{\rm flare}$ and (d) total power of flares $P_{\rm flare}$ versus relative mass of the convective zone $M_{\rm CZ}/M_{\rm star}$ for different spectral types that are indicated with colors.}
	\label{Fig11}
\end{figure}

To do this sort of analysis, first we should compute the mass of the convective zone $M_{\rm CZ}$ for the SGB. We used figure 8 of \cite{duNascimento03} and the fit of it to extract the relationship between the relative mass of the convective zone ($M_{\rm CZ}/M_{\rm star}$) and effective temperature of the stars. For the temperature range of our sample, exponential function can fit well to their data (figure 8 of \cite{duNascimento03} and also figure 4 of \cite{duNascimento2000}). Figure \ref{Fig10} shows the $M_{\rm CZ}/M_{\rm star}$ ratio versus logarithm of effective temperature for our targets. We use Figure \ref{Fig10} to plot the magnetic activity indicators as a function of normalized mass of the convective zone in figure \ref{Fig11}. Autocorrelation index $i_{AC}$ , effective range of light curve fluctuations $R_{eff}$, time occupation ratio of flares $R_{\rm flare}$ and total power of flares $P_{\rm flare}$ have been plotted in panel (a), (b), (c) and (d) of figure \ref{Fig11}, respectively. Gray dashed lines are least squares linear fits to the data. We do not show averaged flux magnitude of flares $M_{\rm flare}$ due to its similarity to $P_{\rm flare}$. It can be seen that magnetic proxies and flare indexes increase by increasing the relative mass of the convective zone. In other words, when the convective zone includes a larger fraction of the stellar mass, the ability of the star to produce magnetic field is higher. For this reason, the stars from earlier types may have lower level of magnetic activity. However, it should be noted that the stars with higher mass have higher luminosity and the detection of flares and other magnetic structures is more difficult. So, their level of magnetic activity may have been underestimated especially in the band pass of the Kepler data that is more sensitive for certain effective temperatures.   

\subsection{Correlation Analysis} \label{subsubsec:corr}  
In order to make a comprehensive comparison, we also do the correlation analysis for the SGB stars, as we did in paper I for the MS stars. The average correlation values of the magnetic and flare parameters with each other between quarters 2-16 are given in table \ref{tab:corr}. Among the nine correlation pairs, only three of them are nonzero.

The first one is correlation of $i_{\rm AC}$-$R_{\rm eff}$ which is around $\sim 0.5$ indicating that larger and darker magnetic structures live longer on the stellar surface. 

The second non-zero correlation value is related to $R_{\rm flare}$-$P_{\rm flare}$ and has positive value; which is obvious because the higher rate of flare frequency, the higher value of total power. The only point is that the correlation value is higher in earlier type stars, for example, the correlation value of F-type stars is four times that of K-type stars. One possible explanation is that in hotter stars, spots are more contrast with respect to the surrounding photosphere \citep{Berdyugina2005} and therefore have stronger magnetic field, this cause a higher power output of magnetic energy when the flare rate increases. 

The last non-zero correlation value belongs to $R_{\rm flare}$-$M_{\rm flare}$, which is negative and has a lower value in the late type stars. This indicates that the flare rate and magnitude are anti-correlated. In other words, a higher number of flares releases energy more regularly and prevents magnetic energy accumulation and the production of large flares. Conversely, large flares are more likely to occur when the number of flares is low.

The general conclusion for correlation analysis of the SGB is the same as the MS. This suggests that all of the governing relations between the magnetic parameters and flares indexes in the MS stars also apply to the subgiant branch.

\floattable 
\begin{deluxetable}{lccccccccccccccc}
	\tablecaption{Average correlation values for pairs of parameters between quarter 2-16 of each spectral type.
		\label{tab:corr}}
	\tablehead{
		&&  \colhead{$i_{\rm AC}$-$R_{\rm eff}$}  & \colhead{$i_{\rm AC}$-$R_{\rm flare}$} & 
		\colhead{$i_{\rm AC}$-$P_{\rm flare}$}  & \colhead{$i_{\rm AC}$-$M_{\rm flare}$} & \colhead{$R_{\rm eff}$-$R_{\rm flare}$}
		&  \colhead{$R_{\rm eff}$-$P_{\rm flare}$}  & \colhead{$R_{\rm eff}$-$M_{\rm flare}$} &
		\colhead{$R_{\rm flare}$-$P_{\rm flare}$} & 	\colhead{$R_{\rm flare}$-$M_{\rm flare}$} \\                           
	}
	\startdata
	 K-type  &&  0.5094  & 0.1412 & -0.0658 & -0.1485 & 0.0905 & 0.0163 & -0.0164 & 0.1182 &  -0.4564 &  \\
	 G-type && 0.5033 & 0.0195 &0.0349 & 0.0130 & 0.0909& 0.1213 & 0.0493 & 0.3835 & -0.3329
	  &  \\
	 F-type  && 0.4831 & -0.0395 & 0.0053 & 0.0465 & 0.0362 &0.0779 & 0.0674& 0.4961& -0.2481&  \\
	\enddata
\end{deluxetable}

\section{Conclusion} \label{sec:conclusion}

In this paper, we studied magnetic activity and flare characteristics of the subgiant stars with spectral type F, G and K and compared the results with the main-sequence stars which were analyzed in paper I. 
The main result of this analysis is the non-dependency of the magnetic proxies and flare indexes versus $P^{-2}R^{-4}$ (alternative to the Rossby number) for the F and G-type subgiants. It means that the rotation rate and size of the stars do no have important effect on the amount of generated magnetic field in dynamo process. 
This is the opposite of what has been seen for the \textbf{MS}, \textbf{in which magnetic activity of the} G, K and M-type stars have clear dependency to the Rossby number and obey saturation behavior (see paper I and also \cite{Yang2019}).
There are no M-type stars on the subgiant branch, and due to the small number of K-type stars, a general conclusion about them is not possible. 
 
However, the positive relation between the stability of the magnetic features and their relative coverage and contrast holds true on the SGB (though with a lower graph slope in comparison with the MS, see figure \ref{Fig4}), bigger and more contrast magnetic structures live longer on the stellar surface.
 
Also the relationship between the relative coverage and contrast of the magnetic features with flare frequency and total power (see panel b of figure \ref{Fig6} and \ref{Fig8}) in the SGB is \textbf{the} same as the MS. Darker and bigger magnetic features produce more flares with higher total power. This positive relationship is even grater around $log(R_{eff})=-2.0$ \textbf{which} is the critical value for all of the flare proxies in the main-sequence and subgiant branch. Beyond this value, flare rate, total power and magnitude increase with a \textbf{larger} slope. It means that when the coverage and contrast of the magnetic features exceeds a certain threshold, the probability of reconnection as well as the amount of accumulated magnetic energy increase significantly.

Magnetic proxies and flare indexes increase by increasing the relative mass of the convective zone $M_{\rm CZ}/M_{\rm star}$. So, when a higher fraction of the stellar mass is convective, the stars produce more magnetic field through dynamo \textbf{process}, and therefore magnetic activity indicators get larger.

The correlation values \textbf{of parameter pairs and their variation} are the same as the MS, so the governing relation between the magnetic structures and flare events are still valid when the stars \textbf{leave} the MS. We obtained three main results from the correlation analysis. First, larger and darker magnetic structures live longer on the stellar surface. Second, contrary to what we expected, the coverage and contrast of the magnetic structures is poorly correlated with the flare rate. This shows that the magnetic structures which produce the prevailing light curve variations are not simultaneous with the flare events. The last result is that flare frequency is anti-correlated with the flare magnitude, because frequent release of magnetic energy prevents excessive accumulation and \textbf{hence} the large flare magnitude.     

The detail study of star's magnetic field behavior and its \textbf{discrepancy} between the MS and SGB may help \textbf{future} researches about the mechanism of magnetic field production either in the convection zone or \textbf{at the} interface with the radiative core. Also statistical analysis of the flare characteristics in different evolutionary state of the stars would guide investigation of the magnetic reconnection and the consequent flare events. Using more accurate and \textbf{broad} data provided by the recent missions will deepen our insight \textbf{on} the magnetic activity of the stars and its impact on the habitable condition of the nearby planets.       



\acknowledgments
We thank the anonymous referee for useful and constructive comments.
This paper includes data collected by the Kepler mission. Funding for the \emph{Kepler} mission is provided by the NASA Science Mission directorate. The \emph{Kepler} data presented in this paper (\emph{Kepler} Data Release 25) were obtained from the Mikulski Archive for Space Telescopes (MAST).

\end{document}